\documentclass[11pt]{article}

\usepackage[utf8]{inputenc}
\usepackage{amssymb,verbatim,epsfig}
\usepackage{amsmath,setspace}
\usepackage[amssymb]{SIunits} 
\usepackage{multicol}
\usepackage{multirow}
\usepackage{lipsum}  

\usepackage[style = numeric-comp, sorting = none, backend=biber,maxbibnames=6,doi=true, eprint=false]{biblatex}
\DeclareSourcemap{
  \maps[datatype=bibtex]{
    \map{
      \step[fieldset=issn, null]
      \step[fieldset=arxiv, null]
    }
  }
}
\usepackage[usenames,dvipsnames]{xcolor} 
\definecolor{MyBlue}{rgb}{0,0,0.45} 
\usepackage{url} 
\usepackage[breaklinks,colorlinks=true, urlcolor=MyBlue, linkcolor=MyBlue, plainpages=false, citecolor=MyBlue,bookmarks=true,bookmarksopen=true,bookmarksnumbered=true]{hyperref}
\usepackage{breakurl} 
\usepackage{graphicx}
\usepackage{longtable}
\usepackage{lineno}
\usepackage[margin=1in]{geometry} 
\usepackage[small,bf]{caption}

\spacing{1} 
\usepackage{appendix}

\title{\bf 
\vspace{-0.75in}
Thermocavitation in gold-coated microchannels for needle-free jet injection} 

\author{Jelle J. Schoppink$^{1*}$ \& Nicolás Rivera Bueno$^1$ \&
David Fernandez Rivas$^1$, \\ 
\small$^1$Mesoscale Chemical Systems group, MESA+ Institute and Faculty of Science and Technology, \\  
\small University  of Twente, P.O. Box 217, 7500 AE Enschede, the Netherlands. \\
\small $^*$ Corresponding author: j.j.schoppink@utwente.nl}

\date{\today}

\begin{document} \maketitle  

    \begin{abstract}
Continuous-wave lasers generated bubbles in microfluidic channels are proposed for applications such as needle-free jet injection due to their small size and affordable price of these lasers. However, water is transparent in the visible and near-IR regime, where the affordable diode lasers operate. Therefore a dye is required for absorption, which is often unwanted in thermocavitation applications such as vaccines or cosmetics. In this work we explore a different mechanism of the absorption of optical energy. The microfluidic channel wall is partially covered with a thin gold layer which absorbs light from a blue laser diode. This surface absorption is compared with the conventional volumetric absorption by a red dye. The results show that this surface absorption can be used to generate bubbles without the requirement of a dye. However, the generated bubbles are smaller and grow slower when compared to the dye-generated bubbles. Furthermore, heat dissipation in the glass channel walls affect the overall efficiency. Finally, degradation of the gold layer over time reduces the reproducibility and limits its lifetime. Further experiments and simulations are proposed to potentially solve these problems and optimize the bubble formation. Our findings can inform the design and operation of microfluidic devices used in phase transition experiments and other cavitation phenomena, such as jet injectors or liquid dispensing for bio-engineering.
\end{abstract}

\vspace{0.1 in}

\vspace{1cm}

\section{Introduction}\label{C4: sec: intro}
There has been an increasing interest in laser-induced needle-free jet injectors in the last 15 years~\cite{Schoppink2022}. These injectors use a laser to generate a fast-growing bubble inside a microfluidic channel containing liquid, which results in the formation of a liquid jet. Early work relied on a nanosecond pulsed laser to instantaneously deliver a large amount of energy~\cite{Han2012,Tagawa2012}. More recently, continuous-wave (CW) lasers have been used to generate the bubble and jet through so-called thermocavitation~\cite{VanderVen2023,Gonzalez-sierra2023}. Using a CW laser has the advantage of a lower price and a much smaller size of the laser, as no active cooling of the laser components is required, which makes them more attractive for several practical purposes, such as needle-free injections for health applications.

However, using a CW laser requires matching the laser wavelength with the liquid, in order to ensure the absorption of optical energy to form the bubble. CW laser diodes in the visible regime (380-700~nm) and the near-infrared (700-1600~nm) are widely available and affordable ($<100\$$), but the absorption of water at these wavelengths is negligible~\cite{Ruru2012}. To ensure sufficient absorption by the liquid itself, there are typically two options: adding a dye to the liquid or using an infrared laser with wavelength larger than 1900~nm. Various dyes have been used in research, such as copper nitrate~\cite{Gonzalez-sierra2023,ZacaMoran2020b,Berrospe-Rodriguez2017} and Direct Red~\cite{OyarteGalvez2019, VanderVen2023}. However, for the application of needle-free injection, dyes are unwanted. The second option is to use a infrared laser (e.g. Thulium, $\lambda$~=~1950~nm) to ensure direct absorption of water~\cite{Schoppink2023-ETFS,Schoppink2023-POF,Schoppink2024}. However, these are expensive ($>1000\$$) and often require a pump laser, making them less portable. For these reasons, the use of a dye or the use of an infrared laser could hinder large scale adoption of jet injectors relying on a CW laser.

Alternatively, a surface coating could absorb the optical energy and heat the liquid, as earlier suggested by Gonzalez-Sierra et al.~\cite{Gonzalez-sierra2023}. Similarly, in Ref.~\cite{Schoppink2022}, we suggested the use of nanoparticles for energy absorption. This has already been used for surface bubble formation with high energy conversion efficiencies up to 12\%~\cite{Detert2020}, although not for the purpose of microjet generation. This heating of plasmonic nanoparticles would allow to use the cheap and small diode lasers in the visible or near infrared without requiring a dye, allowing for a large range of injectable liquids. However, the fabrication of such nanoparticle covered surfaces requires several nanofabrication steps~\cite{Wang2017}, and reproducibly covering a glass surface inside a microfluidic channel is complex.

In this manuscript, we investigate this potential of using a thin gold layer to absorb the optical energy and generate bubbles and jets. We show preliminary results of a comparison between two different gold layer thicknesses for surface absorption of laser energy and the traditional setup which relies on a red dye for volumetric absorption by the liquid itself. 

\section{Experimental methods}\label{C4: sec: methods}
\begin{figure}[b!]
	\includegraphics[width=\linewidth]{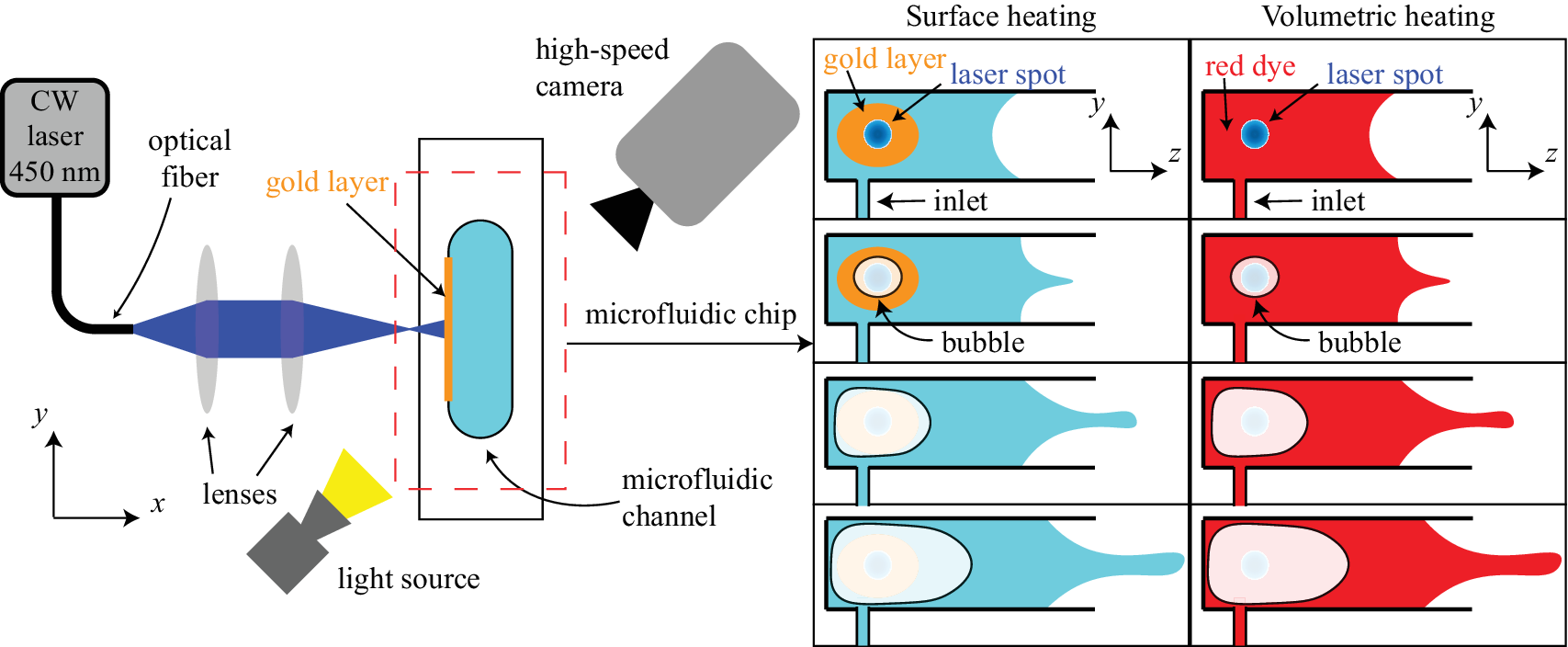}
	\centering
	\captionsetup{width=\linewidth}
	\caption{Schematic of the experimental setup consisting of a blue laser, a microfluidic chip and an high speed camera with illumination. A close-up schematic of the microfluidic chip is shown on the first column on the right. It contains a single rectangular channel, of which one side is partially coated with a gold nanolayer. The laser beam is aligned onto this gold layer, heating up the water inside the channel, resulting in the formation of a bubble and jet. This 'surface heating' is compared to the same experiment with a microfluidic chip without the metallic layer, where a dyed liquid is used for increased absorption, resulting in 'volumetric heating'.}
	\label{C4: fig: ExperimentalSetup}
\end{figure} 

Figure~\ref{C4: fig: ExperimentalSetup} shows the experimental setup, which consists of a blue diode laser focused onto a microfluidic chip that is partially coated with a gold layer. The gold layer absorbs the optical energy and heats up the water, resulting in the formation of a vapor bubble and a microfluidic jet.

The blue laser (Roithner Lasertechnik GmbH, $\lambda $~=~450~nm, P~$\leq$~3.5W) is coupled into a multimode fiber (105~µm core diameter, Thorlabs FG105LCA). It is collimated again at the fiber output before focused onto the microfluidic chip using a 10x objective (Olympus PLN 10x). The use of a multimode fiber results in a Gaussian laser beam, in contrast to earlier work with such a laser diode, where the beam shape resembled an ellipse~\cite{OyarteGalvez2019,OyarteGalvez2020}. During all experiments, the beam radius with intensity 1/e$^2$ ($\approx$ 0.14) at the gold layer or glass interface is 70$\pm$2~µm (see Figure~SI 1 in the supplementary materials) and the laser power is varied between 300 and 1200~mW as it was found to generate a range of different bubble sizes.

The microfluidic chip is made in-house in the MESA+ cleanroom at the University of Twente and consists of two borosilicate glass wafers of 500~µm thickness bonded together. First, the microfluidic channel is wet-etched into the glass wafers. Second, one wafer is partially coated with a tantalum (d = 15 nm) and a gold layer (d = 45 or 90 nm). The tantalum ensures better adhesion of the gold onto the glass. Third, the two glass wafers are bonded together, and the individual chips are diced. The resulting microfluidic channels are 2000~µm long, 400~µm high and 100~µm deep. The gold layer has a surface area of 360 µm by 280 µm. During the experiment, the chips are partially filled with Milli-Q water.

The gold layers are still partially transparent, such that 40\% of the light is transmitted. This allows for visualization of the bubble on the metallic layer as well. The experiments are compared with a Allura Red AC dye, with a concentration of 5 mM, to ensure the same transparency.

\begin{figure}[b!]
	\begin{minipage}[b]{0.5\linewidth}
		
		\centering
		\includegraphics[width=1\textwidth]{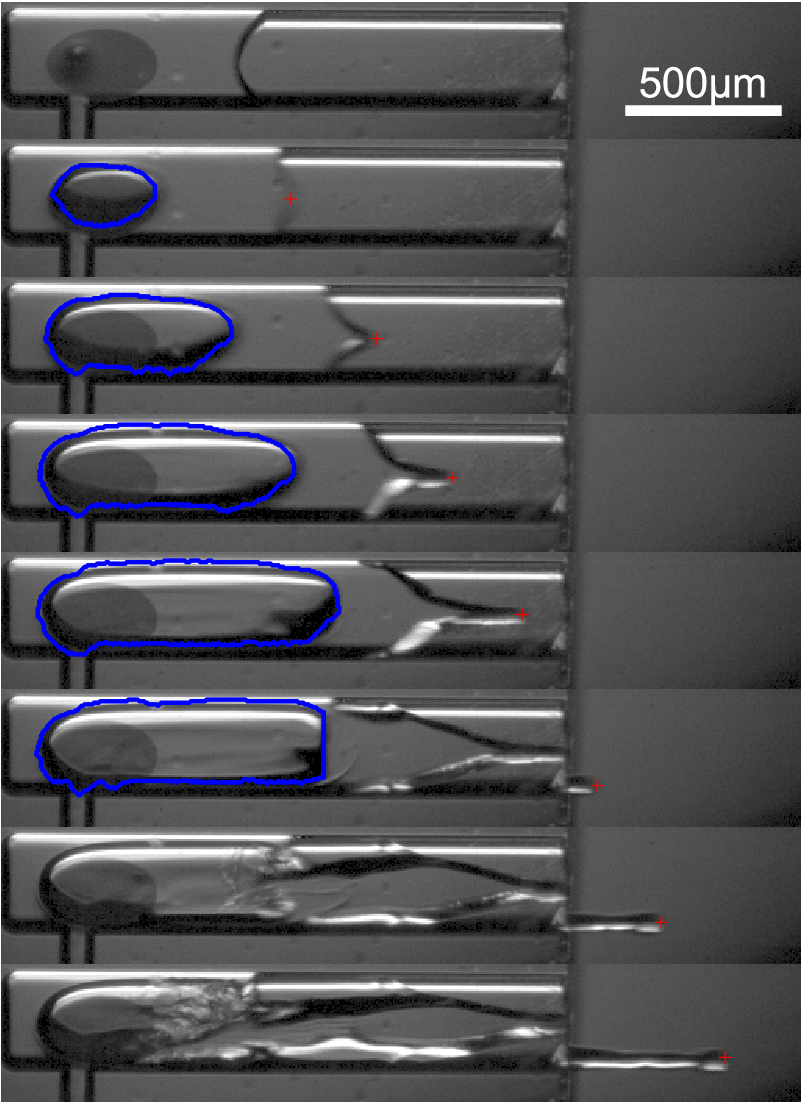}
	\end{minipage}
	\begin{minipage}[b]{0.5\linewidth}
		\centering
		\includegraphics[width=0.9\textwidth]{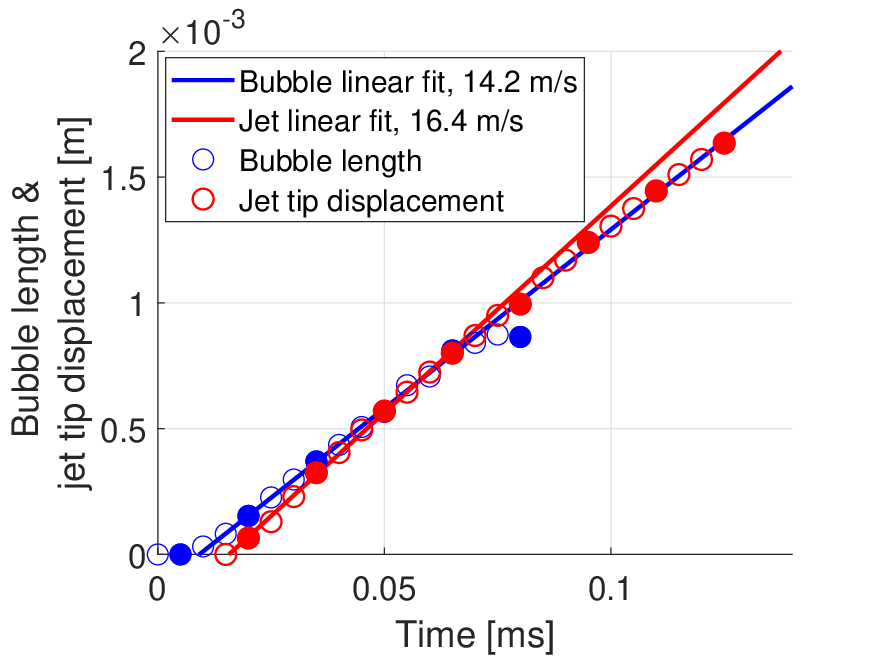}
		\caption{\textbf{Left:} Experimental images of the bubble and the jet with interframe time of 15 µs. Tracked bubble contours are plotted in blue and jet tip position with red '+'. \textbf{Right:} Bubble length (blue) and jet tip displacement (red) over time. The filled markers correspond to the 8 images on the left. A linear fit of the first five data points is performed to calculate the velocities, 14.2 and 16.4 m/s for the bubble and jet respectively. }
		\label{C4: fig: exp images and fits}
	\end{minipage}
\end{figure}

A Photron NOVA SA-X2 high-speed camera was used in combination with a Navitar 12x zoom lens system and a Sugarcube Ultra light source for visualization of the bubble and jet dynamics. Due to the geometrical complexity of the setup and positioning all components, the camera captures images at a 34\textdegree angle, in the $-x,-y$-direction. The camera is used at a framerate of 200k fps, a resolution of 512*88 pixels with a size of 5 by 6~µm. Figure~\ref{C4: fig: exp images and fits} shows eight typical images during the bubble and jet formation. The images are analyzed with a custom-made MATLAB algorithm, which tracks the bubble contours and the jet tip as shown in red. The bubble and jet velocities are calculated by a linear fit of the first five data points of the bubble length (area divided by channel height) and jet tip position respectively.

\section{Results}\label{C4: sec: results} 
In this section, preliminary results and observations of the comparison between surface and volumetric heating will be detailed. A discussion of these results will follow in the subsequent section.

	\begin{figure}[t!]
		
		\begin{minipage}[t]{0.5\linewidth}
			\centering
			\includegraphics[width=\textwidth]{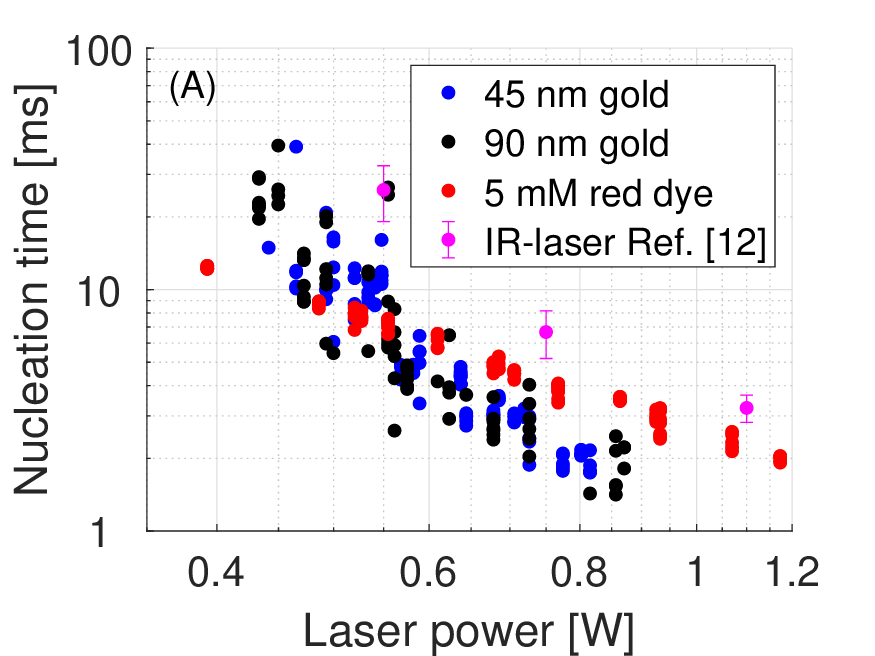}
		\end{minipage}
		\begin{minipage}[t]{0.5\linewidth}
			\centering
			\includegraphics[width=\textwidth]{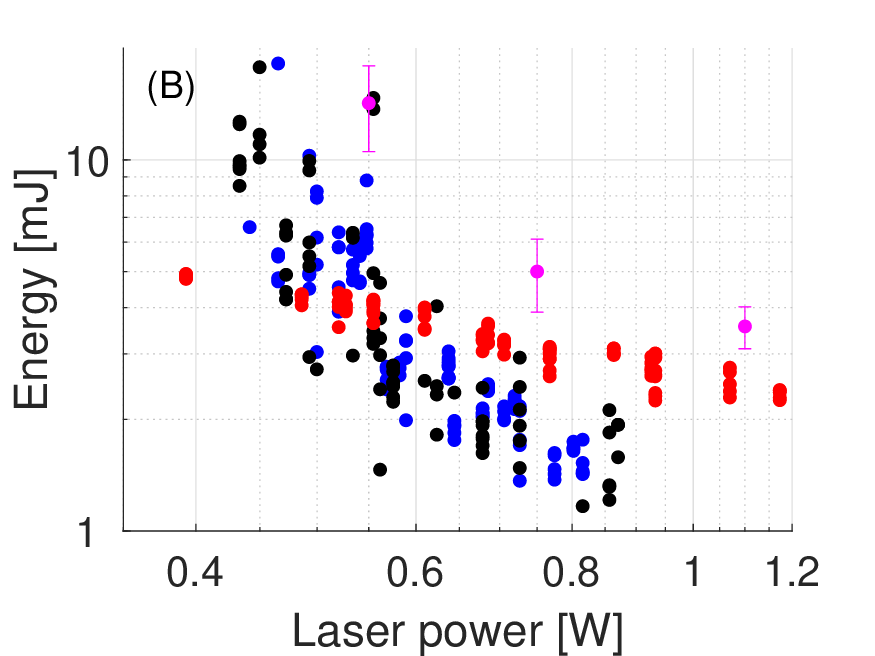}
		\end{minipage}
		\caption{Nucleation time (A) and resulting delivered optical energy to the liquid (B) for a range of laser powers, for the 45~nm layer (blue), 90~nm layer (black) and red dye (red). Each data point represents an experiment creating an individual bubble. The magenta data points are taken from IR-laser experiments of Ref.~\cite{Schoppink2024} with equal beam size, where the error bars indicates the standard deviation.}
		\label{C4: fig: power vs nucl time and energy}
	\end{figure}

Figure~\ref{C4: fig: power vs nucl time and energy}A shows the nucleation times for a range of laser powers. Although the typical nucleation times are both in the order of 1-10 ms, there are some differences between the data of the gold layer and the red dye. First, the slope for the data points of the gold layer is steeper, which indicate that there is a larger influence of laser power. Secondly, while the nucleation times for the red dye are quite reproducible (STD = 5\%), for the gold layers there is a large variation in nucleation time (STD = 15 and 24\% for the 45 and 90~nm gold layers, respectively), indicating another unknown influence. When comparing between the two gold deposited layers, there is no significant difference found, a possible result of the equal transparencies of the gold films of approximately $40\pm2\%$. The Figure also contains the data of the experiments with equal beam size (IR-laser) of Ref.~\cite{Schoppink2024}. The nucleation times in that study are slightly larger compared to the current study. Also, compared to the dye, the slope of these data points is steeper.

The range of delivered energy to the metallic layer is larger compared to the red dye and earlier experiments, which is a direct result of the larger range of nucleation times, see Figure~\ref{C4: fig: power vs nucl time and energy}B. Therefore, for volumetric heating of the dye, the range of energies in this configuration is 2-5 mJ, and controlled through the laser power. For the surface heating, this range is larger, from 1-20 mJ. However, there is a larger variation due to the variability in nucleation times, indicating a reduced reproducibility. 

\begin{figure}[t!]
	\begin{minipage}[t]{0.5\linewidth}
		\centering
		\includegraphics[width=\textwidth]{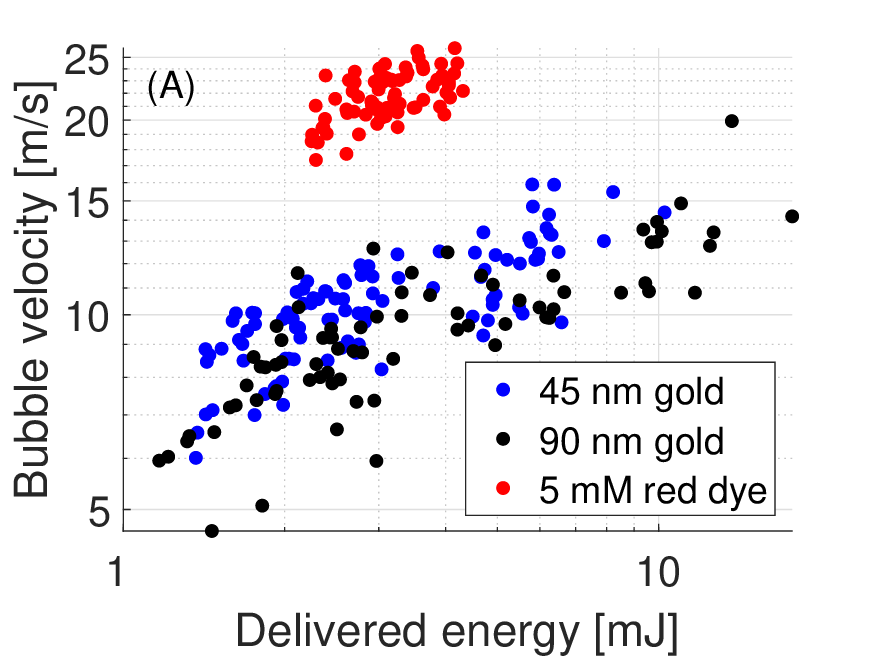}
	\end{minipage}
	\begin{minipage}[t]{0.5\linewidth}
		\centering
		\includegraphics[width=\textwidth]{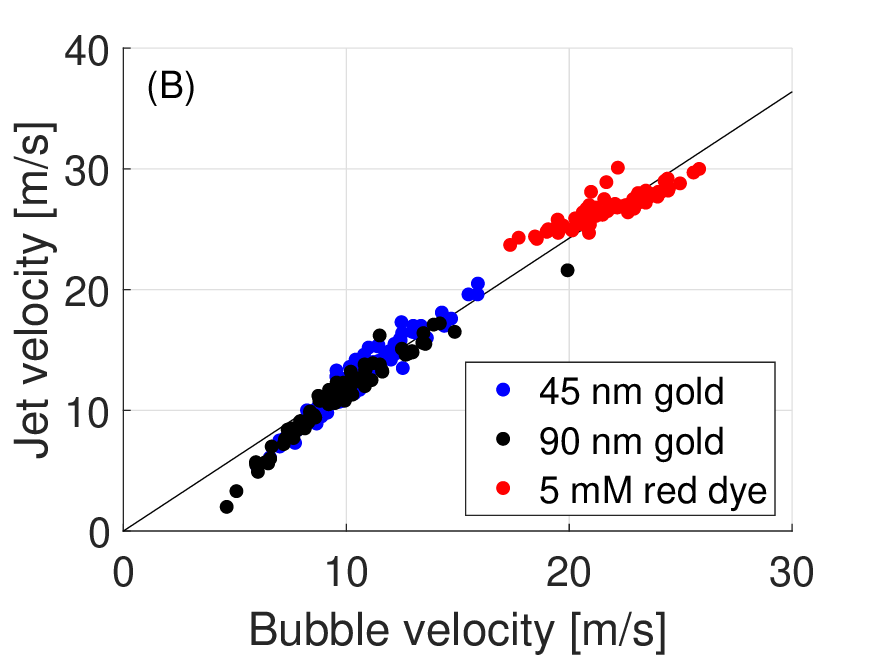}
	\end{minipage}
	\caption{(A) Bubble growth velocities vs. delivered laser energy at nucleation. (B) Jet velocities vs. bubble velocities. In both figures represents each data point an experiment creating an individual bubble and jet. The black line in (B) indicates the best linear fit $y = a \times x$, with $a$ = 1.213.}
	\label{C4: fig: bubble and jet dynamics}
\end{figure}

The bubble velocities are plotted versus the energy delivered to the liquid in Figure~\ref{C4: fig: bubble and jet dynamics}A. Despite the large variation in bubble velocities for individual experiments, there is a positive trend observed between the bubble velocity and optical energy for all cases. For both gold layers, the bubble velocities are in the range of 5-15 m/s, which is smaller than the velocities of 20-25 m/s for the bubbles generated by volumetric heating. 

Figure~\ref{C4: fig: bubble and jet dynamics}B shows the jet velocities versus the bubble velocities. We observed a proportional relation between the jet and bubble velocity for all experiments. As the bubbles generated by volumetric heating of the red dye grow faster compared to the surface heating of the gold layer, the resulting jets are also faster due to increased inertia. Although the jets are faster, the red data points follow the same trend and lie on the same linear curve, (see black line of $y = a \times x$, with $a$ = 1.213), which indicates a similar relation between the bubble and jet velocities.

\begin{figure}[t!]
	\includegraphics[width=0.95\linewidth]{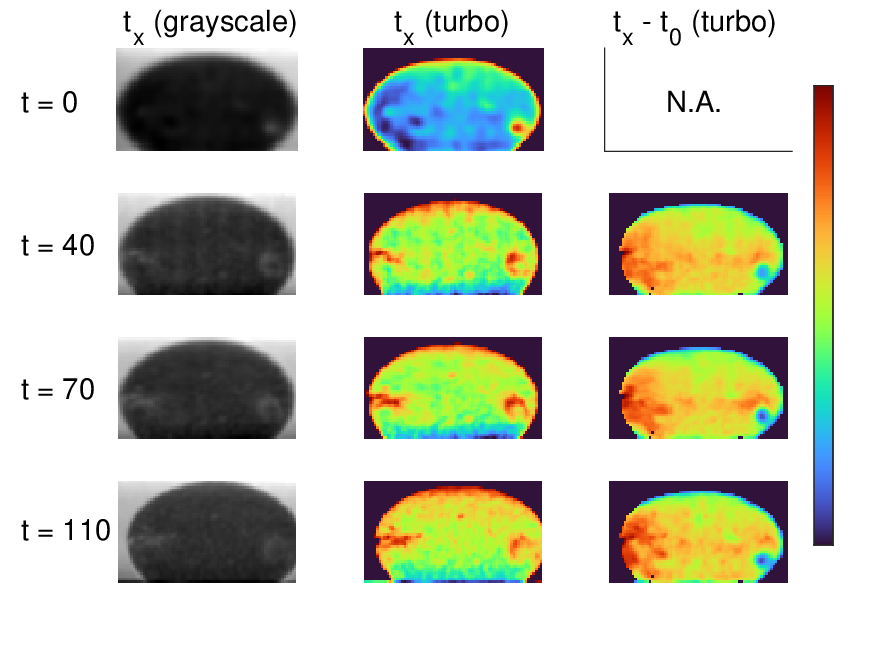}
	\centering
	\captionsetup{width=\linewidth}
	\caption{Images of the 45~nm gold layer over time, indicating layer degradation. \textbf{Left:} Grayscale images at four different instants, where the number on the left indicates the the number of bubbles created prior to this image. \textbf{Middle:} Images in `turbo' colormap to increase contrast (colorbar shown on the right, where red indicates a larger intensity and thus transparency of the layer). \textbf{Right:} The image of the layer at t = t$_x$, substracted by the image at t$_0$. This image shows the increase in degradation on the left and in the center over time.}
	\label{C4: fig: Degradation}
\end{figure}

The 45~nm gold layer changed over time, see Figure~\ref{C4: fig: Degradation} which shows images of the gold layer after specific numbers of generated bubbles and jets. The images on the right of this Figure are the difference between the image after certain amount of bubbles and jets compared to the initial layer. The red regions in the images indicate increased transparency, and therefore degradation. It is clear that on the left the layer degrades at the position where the bubble nucleates. However, degradation is visible also at other positions along the horizontal center line, as indicated by the red regions. 

\section{Discussion}
\subsection{Absorption and nucleation time}
Surprisingly, the transmittance and nucleation times for both gold layers is similar. Initially it was hypothesized that they would be different, as a thicker layer would be less transparent and absorb more light. On the other hand, plasmonic behavior may play an additional role to normal light absorption. Although optical absorption by the tantalum layer could explain similar transparencies, this is unlikely, as tantalum films are transparent in the visible regime and only absorbs in the UV~\cite{Hirpara2022,Graven1961}. Further experiments with a larger range of gold layer thicknesses, as well as only a tantalum layer may provide more information. Also, uv-vis spectroscopy measurements, or irradiation at different angles could conclude whether plasmonic behavior plays a role~\cite{Brockman2000}.

When compared to the dye, we found that the nucleation times are in the same order of magnitude (see Figure~\ref{C4: fig: power vs nucl time and energy}A), but the dependency of laser power is different and there is a larger variation in nucleation times. The difference in heating mechanism (surface vs. volumetric) could explain the difference in laser power dependence. For the surface heating of the gold layer, the temperature gradients in x-direction (along the laser beam) are much larger compared to the volumetric heating of the dye, as the absorption is only at the surface and thus very localized. The timescale $\tau$ on which heat dissipation plays a role is calculated as

\begin{equation}
	\tau = \frac{\delta^2}{4\kappa},
\end{equation}
where $\delta$ is the length scale of thermal diffusion and $\kappa$ the thermal diffusivity (0.14 mm$^{2}$ for water). Due to the more localized energy absorption and larger temperature gradient, $\delta$ is smaller for the gold layer, for which reason heat dissipation already plays a significant role on shorter timescales. Then, as reducing the laser power increases the nucleation time, there is even more heat dissipation and thus nucleation time is further increased, creating a compound effect. For the volumetric heating, the temperature gradient in the x-direction is smaller, and therefore the relevant length and timescales of heat dissipation are larger. 

Similarly, for the earlier experiments of Ref.~\cite{Schoppink2024}, the nucleation times are larger, and the influence of laser power is also larger compared to the experiments with the dye. Although the absorption coefficients of the dye at 450~nm and water at 1950~nm are similar at room temperature, the absorption coefficient of water decreases with increasing temperature~\cite{Schoppink2024}. Therefore, the average absorption coefficient is lower, resulting in larger nucleation times. Also, the curved interface in Ref.~\cite{Schoppink2024} results in more heat dissipation into the glass, which further explains the larger nucleation times.

\begin{figure}[b!]
	\centering
	\includegraphics[width=0.5\textwidth]{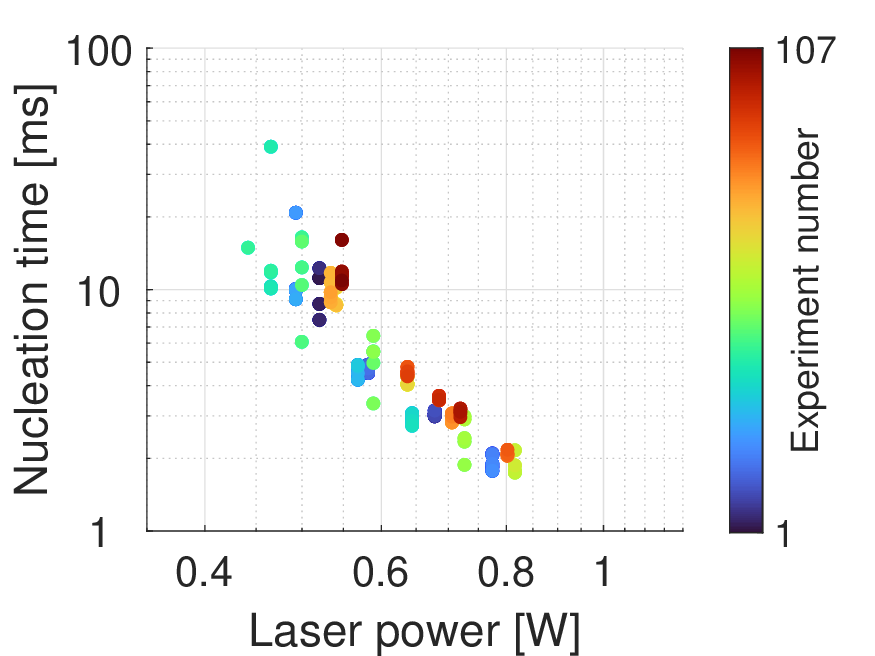}
	\caption{Nucleation times versus laser power for the bubbles generated on the 45~nm gold layer. Each symbol indicates an individual experiment, where the first experiment is shown in blue, and the last ($107^{\textrm{th}}$) in red (see colorbar). It can be observed that later experiments typically have a longer nucleation time.}
	\label{C4: fig: nucleation time degradation}
\end{figure}

For the metallic layer, this increased heat dissipation compared to the dye experiments allows for a larger control over the delivered energy on the range of laser powers, as delivered energies span over a ratio of 20 (1-20~mJ), whereas for the dye this ratio between the maximum and minimum energy is only 2.5 (2-5~mJ). However, the variation in nucleation times for a constant laser power affect the reproducibility and reduce the controllability of bubble and jet dynamics. The degradation of the layer appears to have an effect on this reproducibility. Figure~\ref{C4: fig: nucleation time degradation} shows the nucleation times of all bubbles created on the 45~nm gold layer. It is clearly visible that for the later generated bubbles, the nucleation time is longer, which is most likely caused by the layer degradation. This layer degradation has several effects on thermocavitation. First of all, an increase in transparency would reduce the absorption of energy and delay nucleation. On the other hand, the layer degradation may increase the roughness, and therefore assist bubble formation by reducing the energy barrier. Furthermore, roughness also increases the surface area, which would increase the heat transfer from the gold layer to the water~\cite{Jones2009}. Overall, due to the delayed nucleation for the later-performed experiments, the increase in transparency appears to be the dominant effect.

\subsection{Bubbles and jet dynamics}
In this study the primary goal is not to generate the fastest bubbles and jets, but to initially compare the two methods of bubble formation. This rectangular channel geometry allows for easier analysis of bubble and jet dynamics compared to the curved channels. The potential inclusion of a taper angle allows for faster jets through the incompressibility of water. Earlier experiments in our group with microfluidic channels with a tapered orifice resulted in faster jets in the range 65-94~m/s~\cite{VanderVen2023,OyarteGalvez2020,Berrospe-Rodriguez2017}. Similarly, González-Sierra et al. recently showed that bubble velocities of 10-25 m/s allowed for jet velocities of 70~m/s with a tapered orifice, sufficient to penetrate agar gels and ex-vivo porcine skin~\cite{Gonzalez-sierra2023}.

Figure~\ref{C4: fig: bubble and jet dynamics}A shows the bubble velocities, which are in the range of 5-17 m/s for the gold layer, and 17-26 for the dye. In both cases, there is an increasing trend visible with delivered energy, similar to earlier work~\cite{Schoppink2023-ETFS,Schoppink2024}. When comparing to the results of Ref.~\cite{Schoppink2024}, the bubble velocities for volumetric absorption are similar. There, for the same delivered energies and filling level, the velocities were 20-30~m/s, compared to 18-25~m/s here. This small difference is most likely caused by the different orientation of the set-up (angle of laser irradiation) and liquid (dye vs water).

Although the velocities of the bubbles generated on the gold layer are significantly smaller compared to the dye, the relation with the jet is similar. Figure~\ref{C4: fig: bubble and jet dynamics}B shows the resulting jet velocities, where the red data points lie on the same linear curve as the gold layer experiments. This indicates that the interaction between the bubble and the jet is similar. Therefore, it is hypothesized that if faster growing bubbles would be generated on the gold layer, they would create similar jets.

The slower bubble growth and velocities can be explained by the difference in (super-)heated volume. To make a numerical comparison between the two types of heating, a simplified numerical simulation was performed using the one-dimensional heat equation. In this simulation, only the liquid close to the wall was heated to mimic the indirect surface heating through the metallic layer. For the volumetric heating, the whole liquid was heated with an exponential decay to mimic the reducing laser irradiance. In both cases heat diffusion in the liquid was included. More details regarding these simulations can be found in Section~SI 2 in the supplementary materials. 
Figure~\ref{C4: fig: volumetric vs surface heating 20 ms} shows the normalized temperature profiles. The volumetric heating results in a much more homogeneous temperature profile compared to the surface heating, where only the liquid close to the surface (x = 0) is heated. Although the nucleation times are comparable to the thermal diffusion time of 18 ms (for a length of 100~µm), the temperature increase at x = 100~µm is only approximately 22\% compared to the increase at x = 0. Due to this inhomogeneous temperature profile, the liquid at the surface may be heated faster, but at the moment of nucleation, the superheated volume is smaller. Although this is a largely simplified simulation, it semi-quantitatively explains the smaller and slower growing bubbles for the metallic layer (surface heating) compared to the dye (volumetric heating). Generating faster growing bubbles requires a larger superheated volume~\cite{Schoppink2024}. Therefore, to generate larger and faster growing bubbles, the metallic layer should have a larger heated surface area, and/or the laser power should be further reduced to increase heat dissipation. For a full quantitative comparison, future simulations should be performed in 3D and include an interfacial thermal resistance between the gold-glass and gold-water interfaces.  These simulations could be used for optimization of channel geometry and laser and liquid parameters to maximize superheated volume and therefore bubble growth.

\begin{figure}
	\centering
	\includegraphics[width=0.5\textwidth]{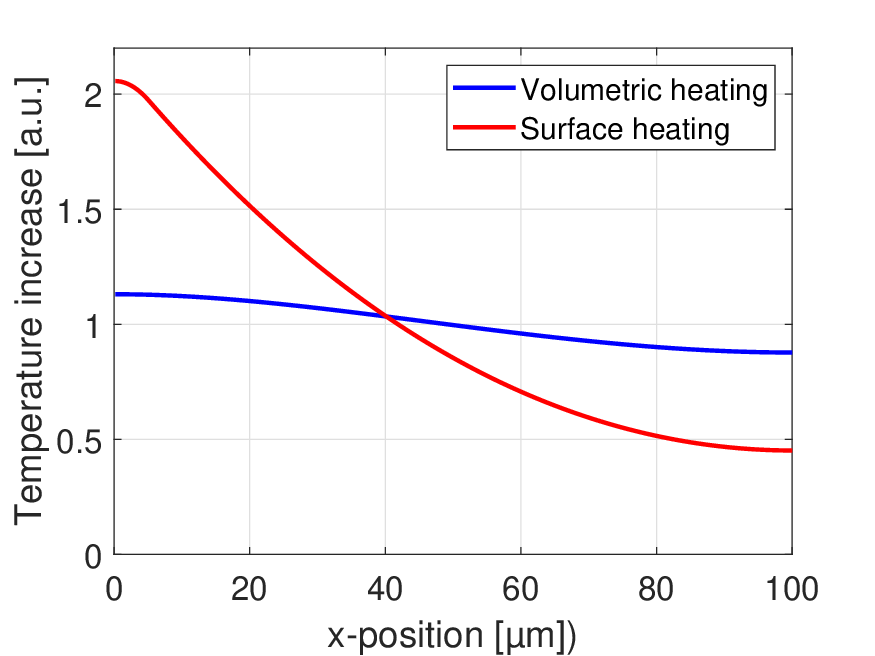}
	\caption{Simulated normalized temperature profile after 20 ms of volumetric (red) and surface heating (blue). The numerical simulations of the one-dimensional heat equation include adding heat at every timestep, mimicking the absorption of optical energy. In the case of volumetric heating, the whole liquid is heated, whereas in the case of surface heating, only the liquid at the surface is heated.}
	\label{C4: fig: volumetric vs surface heating 20 ms}
\end{figure}

\subsection{Energy efficiency}
As mentioned in the previous subsection, the bubbles generated on the metallic layer are smaller and grow slower, which is explained by reduced volume that is heated. However, the combination of similar energies but slower bubbles indicate a lower efficiency. If only a small liquid volume is heated, nucleation should happen earlier at which point the delivered energy is smaller, assuming same nucleation temperature. One explanation is that part of the absorbed energy in the metallic layer is also dissipated into the glass. The thermal diffusivity of the borosilicate glass is 0.64~mm$^{2}$/s~\cite{Schott_Thermal}, which is approximately four times as high as water (0.14~mm$^{2}$/s). Therefore, a nonnegligible amount of the heat absorbed by the gold layer will dissipate into the glass walls, which reduces the efficiency. For the volumetric absorption, dissipation into the glass has a lesser impact, as the energy is absorbed within the liquid, of which most is not in direct contact with the glass. The larger thermal diffusivity of glass results in the fact that the glass is heated up approximately twice as fast (see Figure~SI 4 in the supplementary materials). This is partially canceled out by the volumetric heat capacity of water, which is twice the value of the glass volumetric heat capacity. Therefore, it can be calculated that approximately 50\% of the energy absorbed by the gold layer dissipates into the glass, and the other half into the water (see Section~SI 2.1 in the supplementary materials for calculation). Overall, optimization of the energy transfer therefore requires the use of a substrate with low thermal diffusivity and volumetric heat capacity. Alternatively, a thermally insulating but light transparent layer between the gold and the glass could reduce these energy losses.

However, this 50\% loss in energy does not fully explain the reduced in efficiency of the gold layer compared to the dye. The expected required delivered energy for bubble velocities on the gold layer in Figure~\ref{C4: fig: bubble and jet dynamics}A are even lower than 50\% when extrapolated from the red dye. This reduced efficiency could furthermore be explained by the reflection of the gold layer. Besides the 40$\%$ transmission, significant reflection of the laser light is observed from the gold layer. From qualitative observation during the experiments this reflection is larger compared to the experiments with the dyed liquid in uncoated channels. Unfortunately, due to the geometrical complexity of the set-up it is difficult to measure the reflection, especially at the 90\textdegree~incident angle. A future design could include a anti-reflective coating on the gold layer to reduce the reflections and increase the efficiency.



\subsection{Layer degradation}
Although the degradation event itself was not directly observed in the high speed videos, it is hypothesized that it is caused by the bubble. The degradation is not caused by the heating itself, as long laser irradiation of the layer in air did not cause any visible damage. Furthermore, repeated laser pulses (N$>$100) shorter than the nucleation time (such that no bubble is formed) did not cause any visible degradation. Therefore, we hypothesize that the bubble must play a role in this degradation. As the damage is not confined to the region of nucleation, it is hypothesized that it is caused by the bubble collapse, which is due to its violent nature known to damage proximate surfaces~\cite{Reuter2022,Rossello2022a}. However, as most damage is found at the nucleation site, the nucleation event may also play a significant role. This can be explained as there are stresses built-up during the heating phase, which are released at nucleation.

For the application of needle-free jet injection, this layer degradation poses an unacceptable risk. These released gold nano-/microparticles could be injected together with the liquid, which is unwanted. Furthermore, this layer degradation is also found to affect the heating phase of subsequent bubble and jet formation, reducing the reproducibility and useful operation time of the microfluidic device.

To reduce the degradation, the gold layer could be covered by a thin protecting layer of titanium, which is more stable than gold~\cite{Shlepakov2020,FernandezRivas2012}. Alternatively, the microfluidic chip could be (partially) made of silicon, with a small surface area made rough to create 'black silicon', which is highly absorbing on a large range of wavelengths~\cite{Fan2021}. Furthermore, the mechanical stability of black silicon can be improved by a thin coating of Al$_{2}$O$_{3}$~\cite{Schmelz2023}. If these substrates are still vulnerable to degradation, natural dyes could be employed to create bubbles~\cite{Afanador-Delgado2020}, but those would limit the range of injectable liquids. Alternatively, a membrane could split the dyed-liquid where the bubble is formed and the non-dyed injected liquid~\cite{Han2010}, but that reduces the jet velocity as there are losses in the energy transfer~\cite{EbrahimiOrimi2020}.

\section{Conclusion}
In this manuscript, a preliminary study compares two methods of thermocavitation: surface heating of a thin gold layer and volumetric heating of a red dye. The gold layer would have the advantage of not requiring a dye, important in thermocavitation applications such as needle-free jet injection. Although the results are preliminary, it is clear that the gold layer can be used to heat the liquid and generate bubbles and jets. However, several disadvantages have been observed. First, surface heating is indirect and only a part of the absorbed energy is transferred to the liquid, which reduces the efficiency. Furthermore, the heating of liquid relies on heat diffusion, which is slow and therefore only a small amount of liquid is heated. This results in smaller superheated volumes and slower bubbles and jets. Longer nucleation times, larger heated areas and/or thinner microfluidic channels could solve this. Finally, over time the layer degraded, which is caused by the nucleation or bubble collapse. This reduces the reproducibility over time, and contaminates the liquid with gold microparticles, both of which would be a large problem for most applications. Future studies should focus on solving these different aspects and improving the bubble formation on the metallic layer.

\section*{Supplementary materials}
The supplementary information contains further details on the laser beam sizes (SI 1) and numerical calculations on the heating phase for volumetric and surface heating (SI 2)

\section*{Acknowledgements}
J.J.S and D.F.R. acknowledge the funding from the European Research Council (ERC) under the European Union’s Horizon 2020 Research and Innovation Programme (Grant Agreement No. 851630). J.J.S. would like to thank Stefan Schlautmann for the fabrication of the microfluidic chips and Frans Segerink and the Optical Sciences group of the University of Twente for their help with the optical set-up.

\section*{Competing interest}
J.J.S. is CTO of FlowBeams and D.F.R. is CScO and co-founder of FlowBeams, which is a spin-off company of the University of Twente on laser-actuated needle-free injection.

\section*{CRediT authorship contribution statement}
\textbf{Jelle Schoppink:} Conceptualization, Methodology, Formal analysis, Investigation, Data Curation, Writing - Original Draft, Visualization
\textbf{Nicolás Rivera Bueno:} Conceptualization, Methodology, Formal analysis, Investigation, Data Curation, Writing - Review \& Editing
\textbf{David Fernandez Rivas:} Conceptualization, Supervision, Project administration, Funding acquisition, Writing - Review \& Editing.

\section*{Data Availability Statement}
The data that support the findings of this study are available from the corresponding  author upon reasonable request

\printbibliography

\end{document}


\maketitle

\begin{supplement}

\section{Laser beam shapes}\label{C4: Suppl: Sec: Beam shapes}

\begin{figure}[b!]
	\centering
	\vspace{3 mm}
	\begin{minipage}[t]{0.3\linewidth}
		\centering
		\includegraphics[width=\textwidth]{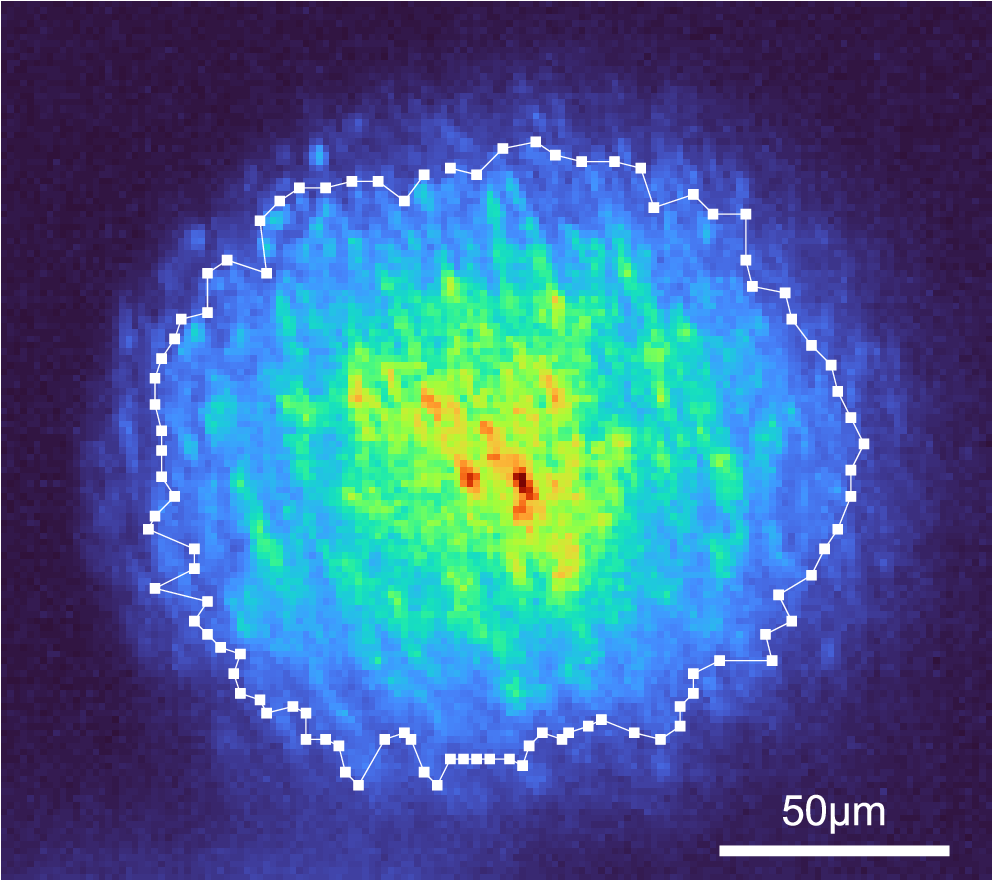}
	\end{minipage}
	\hspace{0.5cm}
	\begin{minipage}[t]{0.3\linewidth}
		\centering
		\includegraphics[width=\textwidth]{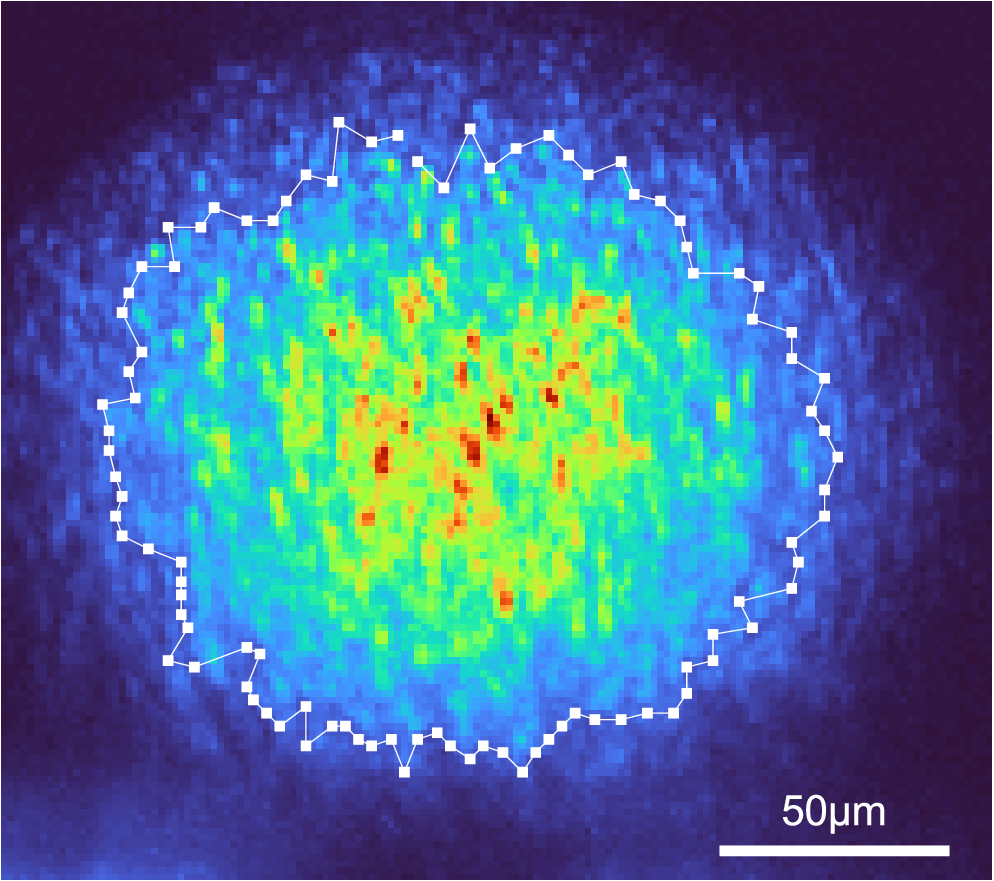}
	\end{minipage}
	\caption{Laser beam shapes on the metallic layer captured by the camera for the 45~nm (left) and 90~nm (right) layer, imaged in colormap `turbo'. The white boxes correspond to the normalized intensity equal to 1/e$^{2}$ ($\approx 0.135$), from which the calculated beam radii are 69 (left) and 68~µm (right).}
	\label{C4: suppl: fig: beam shapes}
\end{figure}
To create a reproducible beam shape for each experiment, the microfluidic chip is positioned using 3-axis stage (Thorlabs Rollerblock) which allows for micrometer accuracy positioning. As the laser beam is diverging, exact positioning is required to ensure the same identical beam size. Prior to each experiment, the beam size is imaged using the camera. 
Figure~\ref{C4: Suppl: Sec: Beam shapes} shows the laser beam shapes on the metallic layer, for the 45 and 90~nm, respectively. These images are taken in the same configuration as the experiments (see Figure~1 in the main manuscript), but without the orange filter front of the camera, which normally blocks the blue laser light to protect it. As the camera is positioned at an angle, the images show the scattered laser light on the metallic layer. In both figures, the beam radius (1/e$^{2}$) is found to be approximately 70~nm. However, after further analysis, it was found that the intensity profiles are slightly different. The right image (90~nm) has a wider region of high intensity (red dots), whereas the high-intensity region in the left image is smaller. As the heat diffusion on the gold layer is much faster compared to the nucleation times, it is hypothesized that this is not significant. Furthermore, assuming the delivered energy is constant, the beam size does not have a significant effect on the bubble dynamics~\cite{Schoppink2024}.

For the glass chip, the imaging of the laser beam is more complex. Due to the lack of the gold layer, there is (almost) no light scattering on the surface. Only minor surface defects result in scattering, but they are less abundant. To ensure the same beam size, the camera is kept at the exact same position when changing to the microfluidic chip without metallic layer. By moving the chip in focus of the camera, it is (approximately) in the same position as the previous chip. For further confirmation of the beam size, imaging of the beam size is still possible, although much noisier. Figure~\ref{C4: suppl: fig: beam shapes_glass} show these shapes, which confirm approximately the same beam diameter on the glass channel.

\begin{figure}[t!]
	\centering
	
	\begin{minipage}[t]{0.3\linewidth}
		\centering
		\includegraphics[width=\textwidth]{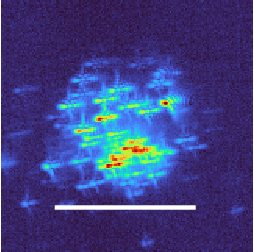}
	\end{minipage}
	\hspace{0.5cm}
	\begin{minipage}[t]{0.3\linewidth}
		\centering
		\includegraphics[width=\textwidth]{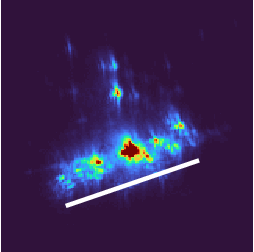}
	\end{minipage}
	\caption{Laser beam shapes on the glass chip interface captured by the camera for the 45~nm-coated channel (left) and uncoated channel (right), imaged in colormap `turbo'. Compared to imaging the beam on the gold layer, visualization is more complex as there is almost no scattering on the smoother glass. Nonetheless, minor defects on the glass do scatter the laser light, providing an estimate of the beam size. The length the white lines in the image correspond to 135 µm and are approximately equal to the beam diameter.}
	\label{C4: suppl: fig: beam shapes_glass}
\end{figure}

\section{Heating phase simulations}\label{C4: Suppl: Sec: Heating phase simulations}
To compare the heating phase of the water through the gold layer and the dye, a very simplified numerical simulation is performed to obtain the temperature profile. The one-dimensional heat equation (see Equation~\ref{C4: suppl: eq: heat equation}) is simulated over time in a custom-made MATLAB script. 

\begin{equation}\label{C4: suppl: eq: heat equation}
	\frac{\delta T(t,x)}{\delta t} = \kappa \frac{\delta^2 T(t,x)}{\delta x^2} + Q(x)
\end{equation}

In this equation, $T$ is the temperature, $t$ indicates time, and $x$ the spatial coordinate. 
This simulation includes heat dissipation in the liquid, calculated from the the liquid thermal diffusivity of water ($\kappa$ = 0.14~mm$^{2}$/s). Furthermore, the temperature is locally increased at every time step to mimic the laser heating, indicated by $Q(x)$ in the equation. 

In the case of volumetric heating, the added temperature $Q_v$ follows an exponential curve according to Lambert-Beer, see equation~\ref{C4: suppl: eq: volumetric heat}
\begin{equation}\label{C4: suppl: eq: volumetric heat}
	Q_v(x) = C_v \times \exp{(-\alpha x)},
\end{equation} 
where $\alpha$ is the absorption coefficient of the dye, which is approximately 90~cm$^{-1}$, such that the ratio of $Q_v$ on the right boundary compared to the left boundary ($Q_v(x = 100~\mu m)$/$Q_v(x=0)$) is approximately 0.4. The constant $C_v$ is included to normalize $Q_v$. The added temperature per unit time for surface heating $Q_s$ can be seen in equation~\ref{C4: suppl: eq: surface heat}, 
\begin{equation}\label{C4: suppl: eq: surface heat}
	  Q_s(x) = \begin{cases} C_s & \mbox{if } 0 \le x \le 5 \mu m \\ \mbox{0} & \mbox{if } x > 5 \mu m \end{cases}
\end{equation} 
where all added temperature is in the region close to the surface, over a length of 5~µm. This would mimic the heated of the region only in close contact with the metallic layer. For larger values of x, the temperature is not increased, as it is not in contact with the metallic layer, and this region is only heated through heat dissipation. The constant $C_s$ is included to ensure that the sum of $Q_s$ and $Q_v$ are equal, such that the average temperatures are the same in both simulations.

The spatial length of the simulation is taken as 100~µm, similar to the channel thickness in the experiment. This is split in 500 grid points, seperated by 200 nm. The heating phase is simulated for a total of 20 ms, similar to typical experimental nucleation times. The simulation includes $10^{6}$ time steps, to ensure convergence of the simulation. Heat dissipation into the boundaries is not taken into account, as it would largely increase the complexity and would most likely affect both simulations equally.

\begin{figure}[t!]
	\centering
	\includegraphics[width=\textwidth]{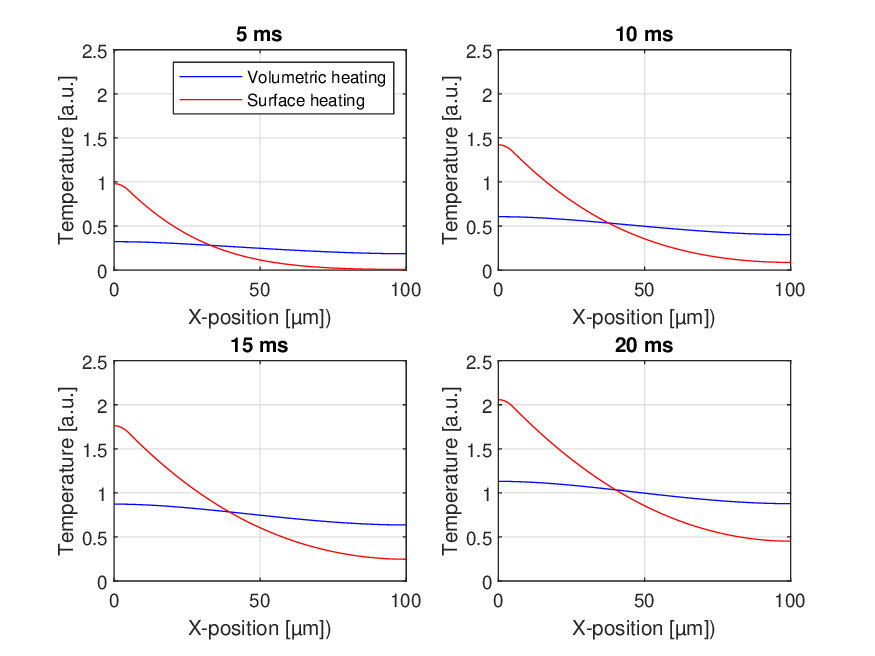}
	\caption{Simulated normalized temperature profiles for volumetric vs surface heating at four different time instants. The numerical simulations of the one-dimensional heat equation include locally increasing temperature at every timestep to mimic the absorption of optical energy. In the case of volumetric heating, the whole liquid is heated, whereas in the case of surface heating, only the liquid at the surface is heated.}
	\label{C4: suppl: fig: volumetric vs surface heating}
\end{figure}

The resulting temperature profiles at four time instants (5, 10, 15 and 20 ms) are shown in Figure~\ref{C4: suppl: fig: volumetric vs surface heating}. In the case of surface heating (red curve), the temperature increase for small values of $x$ (close to the metallic layer) is high, whereas the temperature increase for large $x$ is close to zero. As the time for thermal diffusion to act over a length of 100~µm is approximately 18 ms ($t = \frac{L^2}{4\kappa}$), the temperature at x = 100~µm only starts increasing at larger times. Therefore, even at 20 ms, the temperature profile is still largely inhomogeneous, such that most heat is localized close to the metallic layer. 

In contrast, for the volumetric heating, the whole liquid is heated, resulting in a more homogeneous temperature profile.
Although the liquid close to the left wall (x=0) is heated faster due to the exponential decay of the laser irradiance (see Equation~\ref{C4: suppl: eq: volumetric heat}), heat dissipation flattens this curve over time.

Although the simulations are largely simplified, it semi-quantitatively shows that in the case of surface heating, the temperature profile is not constant. On the experimental timescales (1-20 ms) only the liquid close to the metal layer is heated. On the other hand, in the case of volumetric heating, the optical energy is already absorbed over a larger length, for which reason a much larger volume of liquid is heated. Therefore, it can be concluded that for volumetric heating, the superheated volume is larger, which generates a faster growing bubble. 

\subsection{Surface heating including glass}\label{C4: Suppl: Subsec: simulations including glass}
As mentioned, for the surface heating, dissipation into the glass cannot be neglected. The thermal diffusivity of glass is $\alpha_g =$ 0.64~mm$^{2}$/s, more than 4 times as high as the one of water ($\alpha_w$ = 0.14~mm$^{2}$/s). When including the glass layer in the simulation, it becomes clear that the glass heats up faster than the water, as can be seen in Figure~\ref{C4: suppl: fig: surface_heating_glass_water}. The ratio between the temperature increase in the glass and the water (grey vs blue area) is equal to 0.68 : 0.32. The origin of the 0.68 lies in the ratio of the thermal diffusivities, where $0.68\approx(\frac{\alpha_g}{\alpha_g+\alpha_w})^2 = (\frac{0.64}{0.78})^2$. The square relation originates from the fact that the temperature gradients are not equal, which is much larger in the water compared to the glass (as can also be seen in Figure~\ref{C4: suppl: fig: surface_heating_glass_water}).

\begin{figure}[b!]
	\centering
	\includegraphics[width=0.5\textwidth]{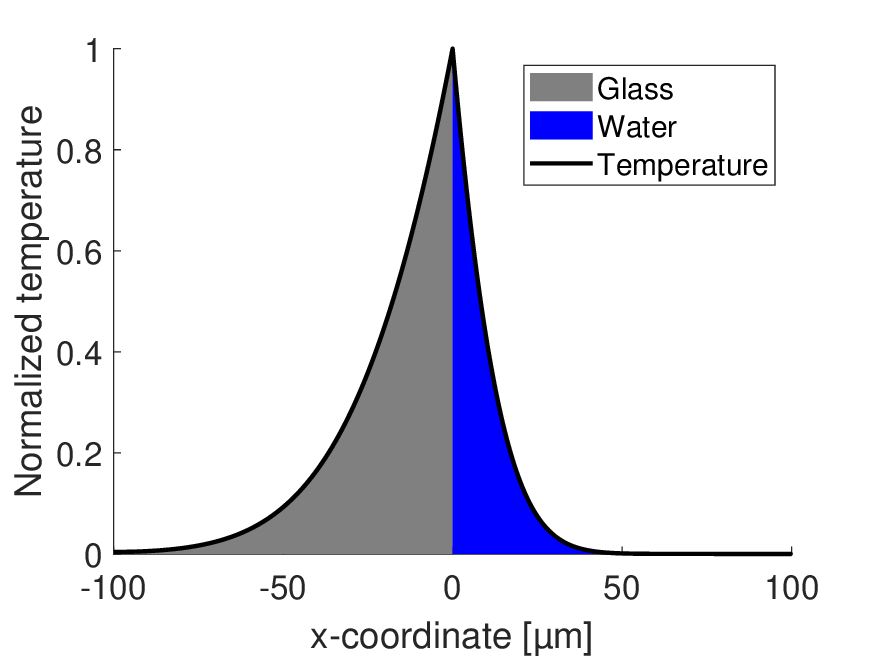}
	\caption{Simulated normalized temperature profiles for surface heating. The numerical simulations of the one-dimensional heat equation include locally increasing temperature at x = 0 at every timestep to mimic the absorption of optical energy. Heat dissipation into the glass x$<0$ and the water x$>0$ is included according to their heat diffusivity of 0.64 and 0.14 mm$^{2}$/s, respectively. The areas under the curve have a ratio of 0.68:0.32}
	\label{C4: suppl: fig: surface_heating_glass_water}
\end{figure}

Although the temperature of the glass increases approximately twice as fast as the water (0.68/0.32$\approx$2), this does not mean that 68$\%$ of the energy is lost in heating up the glass. The volumetric heat capacity of glass is lower than the one of water. For the borosilicate glass, the specific heat capacity is 0.83~kJ/(kg~K)~\cite{Schott_Thermal}, and the density 2230~kg/m$^{3}$~\cite{Schott_Mechanical}, resulting in a volumetric heat capacity of 1.85~MJ/(K m$^3$), approximately 44$\%$ of the value of water (4.18~MJ/(K m$^3$)). Multiplying these values result in a energy ratio of 0.68*1.85 : 0.32 * 4.18 = 1.26 : 1.33, or approximately 1 to 1. This means that half the absorbed energy dissipates into the glass and the other half into the water. It is important to mention that this calculation does not include the actual heat transfer across the interfaces of the gold to the glass and water with an interfacial thermal resistance, only the heat dissipation in the glass and water. For the glass, this actually includes two interfaces, first from the gold to the tantalum layer and then from the tantalum layer to the glass. Nonetheless, this calculation gives a rough estimation of the energy dissipation into the glass.

\printbibliography
\end{supplement}